\documentclass[final,5p,times,twocolumn]{elsarticle}

\usepackage{graphicx}
\usepackage{cases}
\usepackage{amssymb}
\usepackage{amsthm}
\usepackage{bm}
\usepackage{here}
\usepackage{color}

\definecolor{mygreen}{rgb}{0.2,0.8,0.2}

\begin{document}

\begin{frontmatter}


\title{Role of  self-loop in cell-cycle network of budding yeast
}





\author[label1]{Shu-ichi Kinoshita}
 \address[label1]{Department of Mathematical Engineering, Faculty of Engeneering, 
Musashino University, 3-3-3 Ariake Koutou-ku, Tokyo, 135-0063Japan, and 
Institute for Advanced Study of Mathematical Sciences (MIMS), Meiji University, 
4-21-1 Nakano, Nakano-ku, Tokyo 164-8525, Japan.}
\author[label2]{Hiroaki S. Yamada}
 \address[label2]{Yamada Physics Research Laboratory,
Aoyama 5-7-14-205, Niigata 950-2002, Japan}

\begin{abstract}
Study of network dynamics is very active area in biological and social sciences. 
However, the relationship between the network structure and the attractors of the dynamics
has not been fully understood yet.
In this study, we numerically investigated the role of degenerate self-loops on the attractors 
and its basin size using 
the budding yeast cell-cycle network model.
In the network, 
all self-loops negatively surpress the node (self-inhibition loops)
and the attractors are only fixed points, i.e. point attractors.
It is found that there is a simple division rule of 
the state space by removing the self-loops 
when the attractors consist only of point attractors.
The point attractor with largest basin size is robust against 
the change of the self-inhibition loop.
Furthermore, some limit cycles of period 2 appear as new attractor 
when a self-activation loop is added to the original network. 
It is also shown that even in that case, 
the point attractor with largest basin size is robust.
\end{abstract}


\end{frontmatter}

\def\ni{\noindent}
\def\nn{\nonumber}
\def\bH{\begin{Huge}}
\def\eH{\end{Huge}}
\def\bL{\begin{Large}}
\def\eL{\end{Large}}
\def\bl{\begin{large}}
\def\el{\end{large}}
\def\beq{\begin{eqnarray}}
\def\eeq{\end{eqnarray}}
\def\beqnn{\begin{eqnarray*}}
\def\eeqnn{\end{eqnarray*}}

\def\bsc{\begin{screen}}
\def\esc{\end{screen}}
\def\bit{\begin{itemize}}
\def\eit{\end{itemize}}
\def\bca{\begin{cases}}
\def\eca{\end{cases}}

\def\bfr{\begin{framed}}
\def\efr{\end{framed}}
\def\bbr{\begin{breakbox}}
\def\ebr{\end{breakbox}}
\def\bqu{\begin{quote}}
\def\equ{\end{quote}}

\def\vsp5{\vspace{5mm}}

\def\eps{\epsilon}
\def\th{\theta}
\def\del{\delta}
\def\omg{\omega}

\def\e{{\rm e}}
\def\exp{{\rm exp}}
\def\arg{{\rm arg}}
\def\Im{{\rm Im}}
\def\Re{{\rm Re}}

\def\sup{\supset}
\def\sub{\subset}
\def\a{\cap}
\def\u{\cup}
\def\bks{\backslash}

\def\ovl{\overline}
\def\unl{\underline}

\def\rar{\rightarrow}
\def\Rar{\Rightarrow}
\def\lar{\leftarrow}
\def\Lar{\Leftarrow}
\def\bar{\leftrightarrow}
\def\Bar{\Leftrightarrow}

\def\pr{\partial}

\def\Bstar{\bL $\star$ \eL}
\def\etath{\eta_{th}}
\def\irrev{{\mathcal R}}
\def\e{{\rm e}}
\def\noise{n}
\def\hatp{\hat{p}}
\def\hatq{\hat{q}}
\def\hatU{\hat{U}}

\def\iset{\mathcal{I}}
\def\fset{\mathcal{F}}
\def\pr{\partial}
\def\traj{\ell}
\def\eps{\epsilon}




\section{Introduction}
Recently,  some networks representing metabolic reactions in the cell and 
gene regulatory responses through transcription factors 
have been elucidated along with progress of experimental systems 
and accumulation technology in the database \cite{database2}．
In addition, researches on characterizing the state of the cells as a complex network 
 utilizing these database have been actively investigated \cite{han07,ferhat09,huang13,tran13}．


Moreover, the deterministic discrete-time dynamics for discrete-state model with 
such network structures have been widely studied on the properties 
of the attractors that represent cellular activity states.
This is because the state space is finite, so it is easy to search the fixed points 
and the periodic solutions using  computer power.
For example, Kauffman {\it et al.} modeled the early cells before differentiation 
with the dynamics of the network, 
and made the type of the attractors correspond to the type of cells 
after the differentiation \cite{kauffman69,iguchi07,kinoshita09,daniels18}．
On the other hand, 
Li {\it et al.} discovered that in the model of the gene regulatory network related to 
the cell-cycle, there is a fixed point with a very large basin size, 
and the transition process to the fixed point corresponds to 
the expression pattern of the gene in each process of the cell-cycle \cite{li04}．
It should be noticed that  in the network of the Kauffman {\it et al.}, 
there is no self-regulating factor (self-loop), but  in the model of Li {\it et al.}
 the existence of the self-loops has influence on the attractors.
Very recently, in other systems such as fission yeast cell cycle and 
mammalian cell cycle, 
the Boolean network models for the regulation have also been studied \cite{yang13,barberis16,luo17}.


In this study, using the same gene regulatory network as Li {\it et al.}
for the budding yeast, 
we clarify the relationship between the fixed points (point attractors)  
with large basin size and 
the presence of the self-loops in the network.
It is found that there is a simple division rule of 
the state space by removing the self-loops,  
and the point attractors with largest basin size (BS) is robust
against the changing the self-loops.
The similar results are obtained for C. {\it elegans} 
early embryonic cell cycles as well \cite{kinoshita18}.


\section{Model}
\label{sec:model}
Here, we give some basic properties of the self-loop in cell-cycle network 
of budding yeast.
Let us take the binary value $\{0, 1\}$ as the state $S_i$ of each node $i$
corresponding to the numbered genes as given in table 1.
The states 1 and 0 correspond to expressed and unexpressed genes, 
respectively and the attractors of the dynamics are associated to cell differentiation.
The effect on the node $i$ from the other node $j (\neq i) $ is defined as
\beq
 B_i =\sum_{j (\neq i)}^Na_{ij} S_j, 
\eeq
where $N$ is the total number of the nodes,  
and $a_{ij}$ denotes matrix element of the weighted adjacency matrix  $A$
representing the interaction between the genes.
We take $a_{ij}=+1$ when the node $j$ positively regulates 
the node $i$ (positive interaction ), and 
$a_{ij}=-1$ when the node $j$ negatively suppresses 
the node $i$ (negative interaction). 

\begin{figure}[htbp]
\begin{center}
\includegraphics[width=8.0cm]{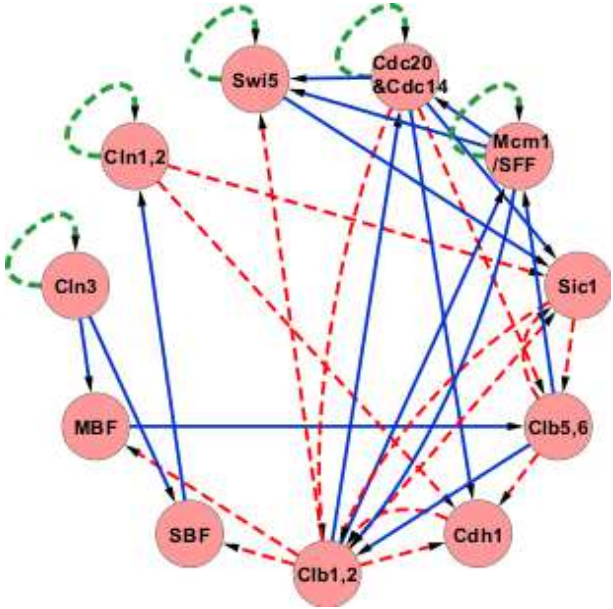}
\caption{
(Color online)
Gene regulatory network of the cell-cycle of budding yeast \cite{li04}．
Each circle represents a protein (cyclin or transcription factor) involved in the gene regulation.
For the links connecting the respective proteins, 
the blue-solid lines represent the effect of the activation control, 
and the red-dashed  lines represent the effect of the suppression control.
In addition, the self-loops by green-dotted lines 
represent the effect (ubiquitin-proteasome system) 
of protein degradation in the absence of external input.
}
\label{fig:yst_net}
\end{center}
\end{figure}

\begin{table*}[t]
\label{table:G0}
\begin{center}
\begin{tabular}{lcccccccccccc} 
 & Cln3 & MBF & SBF & Cln1 & Cdh1 & Swi5 & Cdc20 & Clb5 & Sic1 & Clb1 &  Mcm1 &  \\ 
 No. & 1 & 2 & 3 &4 & 5 & 6 &7 & 8 & 9 & 10 & 11 & BS \\ 
 & $\circ$ & & & $\circ$ &  & $\circ$ & $\circ$ & & & & $\circ$ &  \\ \hline
 $A_1^{(0)}$ & 0  & 0 & 0 & 0 & 1& 0 & 0 & 0 & 1 & 0 & 0 & 1764 \\
 $A_2^{(0)}$ & 0  & 0 & 1 & 1 & 0& 0 & 0 & 0 & 0 & 0 & 0 & 151 \\
 $A_3^{(0)}$ & 0  & 1 & 0 & 0 & 1& 0 & 0 & 0 & 1 & 0 & 0 & 109 \\
 $A_4^{(0)}$ & 0  & 0 & 0 & 0 & 0& 0 & 0 & 0 & 1 & 0 & 0 & 9 \\
 $A_5^{(0)}$ & 0  & 0 & 0 & 0 & 0& 0 & 0 & 0 & 0 & 0 & 0 & 7 \\
 $A_6^{(0)}$ & 0  & 1 & 0 & 0 & 0& 0 & 0 & 0 & 1 & 0 & 0 & 7 \\
 $A_7^{(0)}$ & 0  & 0 & 0 & 0 & 1& 0 & 0 & 0 & 0 & 0 & 0 & 1 \\
\end{tabular}
\end{center}
\caption{Seven attractors in the original gene regulatory network.
(All are point attractors.)
The third line shows that there is a degenerate self-loop when mark $\circ$ is present
in the node.
In the decimal notation, each attractor is displayed as, 
$A_1^{(0)}=68$, $A_2^{(0)}=384$,
 $A_3^{(0)}=580$, $A_4^{(0)}=4$, $A_5^{(0)}=0$, $A_6^{(0)}=516$,
 $A_7^{(0)}=64$.
The last column (BS) represents the basin size of the attractors.
Note that Cln 1 represents Cln 1, 2, Clb 5 represents Clb 5, 6, 
and Clb 1 represents Clb 1, 2.
}
%
\label{table:yst_atr}
\end{table*}

The node without the self-loop, i.e. $a_{ii}=0$, follows a threshold dynamics 
from discrete time $t$ to $t+1$ ($t \in {\bf N}$) 
by using the parallel updating scheme as follows:
\beq
S_i(t+1) = \Biggl( 
   \begin{array}{cc}
   0  & (B_i(t) <\theta_i)   \\
 1 & (B_i(t) > \theta_i) \\
S_i(t) & (B_i(t) = \theta_i), 
\end{array}
\label{eq:rule-1}
\eeq
where $\theta_i$ denotes the threshold value of the node $i$.
Also, if the self-loop acts inactively when $B_i(t)=\theta_i$,
the effect of the protein degradation called "degeneration", 
which is distinguished from a simple inhibition effect, is given as follows;
\begin{equation}
  S_i(t+1) = 
\Biggl( 
   \begin{array}{cc}
    0 & (B_i(t)=\theta_i, a_{ii}=-1) \\
    1 & (B_i(t)=\theta_i, a_{ii}=+1) .
\end{array}
\label{eq:rule-2}
\end{equation}

The budding yeast cell-cycle network model (denoted by $G^{(0)}$) 
by Li {\it et al.}  is a special one in a sense that all nodes 
of the existing self-loops are given as $a_{ii}=-1$.
The network is  shown in Figure 1.
We take the values $\theta_i=0$ for all $i$ in this report.
Each regulatory factor is represented by each numbered 
node ($i = 1, 2, ..., 11$), and the activation effect ($a_{ij}=+1$) and 
suppression effect ($a_{ij}=-1$) are indicated by solid and dashed arrows between the nodes.
There are self-degeneration loops on the 5 nodes, Cln3, Cln1-2, Swi5, Cbe/Cdc14, Mcm1/SFF.
Note that this rule is the same as that of Refs. \cite{tran13} and \cite{li04},  
but it differs from that of \cite{goles13}.
In this network, the total state number is $W=2^{11}=2048$, and 
all steady states are seven point attractors by numbering as 
${\bf A}^{(0)}=\{ A^{(0)}_1,A^{(0)}_2,A^{(0)}_3,A^{(0)}_4,A^{(0)}_5,A^{(0)}_6,A^{(0)}_7 \}$.
The state of the point attractor with the largest basin size among these is 
$A^{(0)}_1=00001000100=68$, where the last number is in decimal. 
According to the study of Li et {\it al.} the following facts are known. 
(i)The attractor with the largest basin $A^{(0)}_1= 68$ corresponds 
to the stationary $G_1$ state in the cell-cycle of the budding yeast.
(ii)When creating the random network model of the same system size $N=11$, 
there is no attractor that corresponds to $A^{(0)}_1$ with a very large basin size.
(iii)One of the trajectories to reach the attractor $A^{(0)}_1$ coincides 
with the trajectory of the actual biological cell-cycle.
(iv)The trajectory corresponding to the biological cell-cycle leading to $A^{(0)}_1$
 is stable against external perturbation.

In addition, the result for the basin size of the attractors in the similar random
 networks with 
same conditions of the structure as the $G^{(0)}$ is given in appendix  \ref{app:random}.
We confirmed  that the occurrence probability of 
 the point attractors with the large basin size ($\geq 1700$) 
is less than 20 percent.
This result is consistent with those in Ref.\cite{huang13}.



These results may be due to all self-loops being degenerate and threshold values 
being zero,  and all the attractors are point attractors only.
Generally, the threshold values are related to adding the active self-loops
at each node.
Note that  for fission yeast cell-cycle model with similar network structure
some limit cycles of period two appear as the attractor because some of 
 the threshold value are not zero \cite{goles13,maria09}. 
Further, notice that when an active self-loop is attached to the node
the state update rule becomes different from those of Tran {\it et al.} 
due to the existence of rule (\ref{eq:rule-2}).


\section{Numerical result}
In this section,  we investigate the effect of the degenerate 
self-loops on the attractors of the original network  $G^{(0)}$.
Therefore, we write the network from which the degenerate self-loop of the $k$th node 
is removed from $G^{(0)}$ as $G^{(-k)} $, and 
 the network with self-activating loop is added to the $m$th node
of $G^{(0)}$ as $G^{(+m)}$.
Here, $k$ selects from the nodes with the self-loop, 
and $m$ selects from the nodes without the self-loop.
The attractor sets are indicated as 
${\bf A}^{(-k)}=\{ A^{(-k)}_1,A^{(-k)}_2,...., ..A^{(-k)}_{n_{-k}} \}$, 
${\bf A}^{(+m)}=\{ A^{(+m)}_1,A^{(+m)}_2,....., A^{(+m)}_{n_{+m}} \}$, and so on, respectively, 
where  $n_{-k}$ and $n_{+m}$ means the number of attractors in the networks
$G^{(-k)} $ and $G^{(+m)}$, respectively.
We can numerically decide the all attractors and the basin size 
because the network has a state space of $2^{11}=2048$ states.


\subsection{Case of removing degenerate self-loop
}
\label{subsec:removing}
In Figure \ref{fig:yst_net} of the original network, degenerate self-loops are included
 in five control factors of Cln3, Cln1-2, Swi5, Cbc20/Cdc14, Mcm1/SFF, 
and the table \ref{table:yst_atr} shows the 7 attractors.
We show  in the table \ref{table:yst_des}
the 11 attractors of  the gene regulatory network $G^{(-1)}$
 which removed the degenerate self-loop of  Cln3 （the first node).

\begin{table*}[t]
\begin{center}
\begin{tabular}{lcccccccccccc}  
No. & 1 & 2 & 3 &4 & 5 & 6 &7 & 8 & 9 & 10 & 11 & BS \\ 
 & & & & $\circ$ &  & $\circ$ & $\circ$ & & & & $\circ$ &   \\ \hline
 $A_1^{(-1)}$ & 1  & 1 & 1 & 1 & 0& 1 & 1 & 1 & 0 & 1 & 1 & 888 \\
 $A_2^{(-1)}$ & 0  & 0 & 0 & 0 & 1& 0 & 0 & 0 & 1 & 0 & 0 & 856 \\
 $A_3^{(-1)}$ & 0  & 0 & 1 & 1 & 0& 0 & 0 & 0 & 0 & 0 & 0 & 87 \\
 $A_4^{(-1)}$ & 1  & 0 & 1 & 1 & 0& 1 & 1 & 0 & 0 & 1 & 1 & 61 \\
 $A_5^{(-1)}$ & 0  & 1 & 0 & 0 & 1& 0 & 0 & 0 & 1 & 0 & 0 & 57 \\
 $A_6^{(-1)}$ & 1  & 1 & 0 & 0 & 0& 1 & 1 & 1 & 0 & 1 & 1 & 52 \\
 $A_7^{(-1)}$ & 1  & 1 & 1 & 1 & 0 & 1 & 1 & 0 & 0 & 1 & 1 & 23 \\
 $A_8^{(-1)}$ & 0  & 0 & 0 & 0 & 0& 0 & 0 & 0 & 1 & 0 & 0 & 9 \\
 $A_9^{(-1)}$ & 0  & 0 & 0 & 0 & 0& 0 & 0 & 0 & 0 & 0 & 0 & 7 \\
 $A_{10}^{(-1)}$ & 0  & 1 & 0 & 0 & 0& 0 & 0 & 0 & 1 & 0 & 0 & 7 \\
 $A_{11}^{(-1)}$ & 0  & 0 & 0 & 0 & 1& 0 & 0 & 0 & 0 & 0 & 0 & 1 \\
\end{tabular}
\end{center}
\caption{
Eleven attractors in the gene regulatory network $G^{(-1)}$ 
which removed the degenerate self-loop of Cln3（the first node）. 
(All are point attractors.)
The last column (BS) represents the basin size of the attractors.
In the decimal notation, each attractor is displayed as, 
 $A_1^{(-1)}=1979$, $A_2^{(-1)}=68$,
 $A_3^{(-1)}=384$, $A_4^{(-1)}=1459$, $A_5^{(-1)}=580$, $A_6^{(-1)}=1595$,
 $A_7^{(-1)}=1971$, $A_8^{(-1)}=4$, $A_9^{(-1)}=0$, $A_{10}^{(-1)}=516$, 
$A_{11}^{(-1)}=64$.
}
%
\label{table:yst_des}
\end{table*}


\begin{figure}[htbp]
\begin{center}
\includegraphics[width=8.0cm]{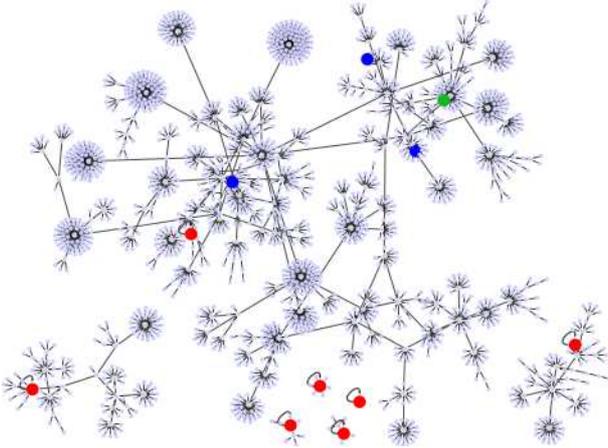}
\caption{
(Color online)
The point attractors and the basin structures of the network $G^{(-1)}$. 
The 7 red circles present the common point attractors to  $G^{(0)}$ and $G^{(-1)}$.
The blue and green circles present  attractors newly added 
by the network becoming $G^{(-1)}$.
The poin attractor with the largest basin of $G^{(-1)}$ is indicated by green circle.
}
\label{fig:Fig2}
\end{center}
\end{figure}

We compare the attractors of the network $G^{(-1)}$ 
with those of $G^{(0)}$.
It is found that 
$A^{(-1)}_{2}=A^{(0)}_{1}$, $A^{(-1)}_{3}=A^{(0)}_{2}$, $A^{(-1)}_{5}=A^{(0)}_{3}$, 
$A^{(-1)}_{8}=A^{(0)}_{4}$, $A^{(-1)}_{9}=A^{(0)}_{5}$, $A^{(-1)}_{10}=A^{(0)}_{6}$, 
$A^{(-1)}_{11}=A^{(0)}_{7}$.
That is, all of the attractor sets ${\bf A}^{(0)}$ of the original network $G^{(0)}$
 is included the attractor set of ${\bf A}^{(-1)}$ of the network $G^{(-1)}$.


Next, we focus on the change of the basin size. 
It follows that the basin size of the attractor $A^{(0)}_1$ with the largest basin size 
 is reduced by the elimination of the degenerate self-loop.
Also, the  basin size of the other attractors are also 
reduced from those of $\bf{A}^{(0)}$.
Figure \ref{fig:Fig2} shows the basin structure of the 2048 initial states 
flowing to the fixed points given in the table \ref{table:yst_des}.
The red circles are the point attractors of $G^{(0)}$, 
and the blue circles indicate the four point attractors newly added 
by the network becoming $G^{(-1)}$.
Obviously, the basin size of the same attractor of $G^{(-1)}$ to those 
of attractor of $G^{(0)}$ is smaller than those 
 of  $G^{(0)}$, and they are caused by branching from the basin of $G^{(0)}$.
Accordingly, it is also easy to understand that all attractors (attractor sets) 
of the original network $G^{(0)}$ are included in the attractor set of $G^{(-1)}$.
The attractor of the large BS of $G^{(0)}$ corresponds to the attractor 
of the relatively large BS of $G^{(-1)}$.





\begin{figure}[htbp]
\begin{center}
\includegraphics[width=8.0cm]{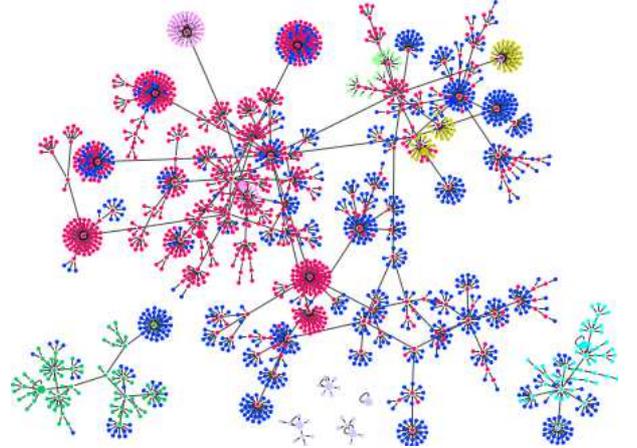}
\caption{
(Color online)
The basin structure of $G^{(0)}$ 
classificated by colors depending on the basins for each attractor 
of $G^{(-1)}$.The states are color-coded so that we can see 
 basins of  the 11 attractors of $G^{(-1)}$.
}
\label{fig:Fig5}
\end{center}
\end{figure}
\begin{figure}[htbp]
\begin{center}
\includegraphics[width=8.0cm]{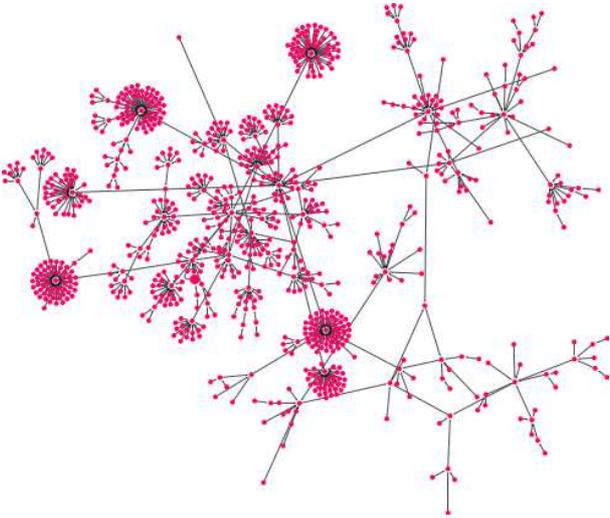}
  \caption{
The basin structure 
that removed the color-coded state other than red in Fig.\ref{fig:Fig5}
from the attractor with  the largest basin of $G^{(0)}$.
}
\label{fig:Fig6}
\end{center}
\end{figure}

In Figure \ref{fig:Fig5}, 
we show the coloring  basin structure of $G^{(0)}$ 
depending on each  basin of the attractors of $G^{(-1)}$.
(Figure \ref{fig:Fig6} shows 
 the one that removed the color-coded state other than red 
from the attractor of the largest basin. )
It is found that the newly appearing attractors of $G^{(-1)}$
are created by connecting the the leaf states to the other leaf states 
in the original gene state in the transition diagram.

Although above results are for the specific case which the degenerate self-loop of Cln3 
has been removed, but also 
 it is found that 
the similar results are also true for the cases removing the other  degenerate self-loops. 
Furthermore, if we apply this rule repeatedly in the process of removing the self-loops, 
we can see that in general the above relations of the attractors and the 
basin size also applies to the relationship before and after removing 
the self-loops.

\subsection{Case of  adding active self-loop
}
\label{subsec:adding}
It is noting that 
in the general network which both the self-regression loops and self-activation loops 
exist, limit cycles can appear as the attractors, as shown in case of the fission yeast.
In networks which the self-activation loop is added to the original network $G^{(0)}$, 
 not only point attractors but also other types of periodic attractors exist.

\begin{figure}[htbp]
\begin{center}
\includegraphics[width=8.0cm]{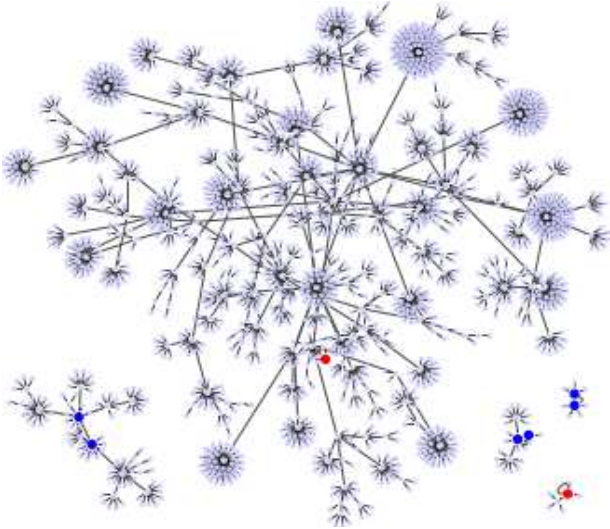}
  \caption{
The attractors and the basin structures of $G^{(+8)}$. 
The 2 red circles present the point attractors.
The 6 blue circles represent the states that belong to the three
limit cycles of period 2 two each, respectively.
}
\label{fig:Fig4}
\end{center}
\end{figure}

As an example, the attractors $\bf{A}^{(+8)}$ of the
 network $G^{(+8)}$ which an active self-loop added to Clb5（the 8th node） 
of the $G^{(0)}$ is given in table \ref{table:yst_add}.
It follows that  the attractors 
$A^{(+8)}_{1}=A^{(0)}_{1}$, $A^{(+8)}_{5}=A^{(0)}_{4}$,  exist also in the network $G^{(0)}$, 
and the limit cycle attractors of period 2,  $A^{(+8)}_2$，$A^{(+8)}_3$，$A^{(+8)}_4$ ,  
are newly emerging as the attractors of the network $G^{(+8)}$.
Also, it follows that 
many attractors of $G^{(0)}$ have disappeared, but the attractor with largest basin size  
 has survived.
The basin structure of  the attractors in the table \ref{table:yst_add}
is shown in Figure \ref{fig:Fig4}.
It is found that the limit cycles 
are constituted by combining the gene states  
with the relatively small basin size.
In such a case  the limit cycles  with large basin size 
do not occur.

These features occur even if the self-activated loop is added to the other nodes
without the self-loop.
Furthermore, the similar phenomena can also be confirmed 
by changing any of the degenerate self-loop of the five nodes  
to the active one.



\begin{table*}[t]
\begin{center}
\begin{tabular}{lcccccccccccc}  
No. & 1 & 2 & 3 &4 & 5 & 6 &7 & 8 & 9 & 10 & 11 & BS \\ 
 &$\circ$ & & & $\circ$ &  & $\circ$ & $\circ$ & $+$ & & & $\circ$ &   \\ \hline
 $A_1^{(+8)}$ & 0 & 0 & 0 & 0 & 1& 0 & 0 & 0 & 1 & 0 & 0 & 1897 \\
 $A_2^{(+8)}(LC_{P2})$ & 0  & 1 & 1 & 1 & 0 & 1 & 0 & 0 & 1 & 0 & 1 & 110 \\
 $ $  & 0  & 1 & 1 & 1 & 0 & 1 & 1 & 1 & 1 & 0 & 0 &  \\
 $A_3^{(+8)}(LC_{P2})$ & 0  & 1 & 0 & 0 & 1 & 1 & 0 & 0 & 1 & 0 & 1 & 25 \\
 $ $& 0  & 1 & 0 & 0 & 1 & 1 & 1 & 1 & 1 & 0 & 0 &  \\
 $A_4^{(+8)}(LC_{P2})$ & 0  & 1 & 0 & 0 & 0& 1 & 0 & 0 & 1 & 0 & 1 & 9 \\
 $ $ & 0  & 1 & 0 & 0 & 0 & 1 & 1 & 1 & 1 & 0 & 0 &  \\
 $A_5^{(+8)}$ & 0  & 0 & 0 & 0 & 0& 0 & 0 & 0 & 1 & 0 & 0 & 7 \\
\end{tabular}
\end{center}
\caption{
Five attractors present in gene regulatory network $G^{(+8)}$ 
which an active self-loop is added to Clb5（the 8th node）.
The three attractors $A_2^{(+8)}$, $A_3^{(+8)}$, $A_4^{(+8)}$ are limit cycle.
$LC_{P2}$ means the limit cycle with the period 2.
The last column (BS) represents the basin size of the attractors.
In the decimal notation, each attractor is displayed as, 
$A_1^{(+8)}=59$, $A_2^{(+8)}=(933,956)$, $A_3^{(+8)}=(613,633)$, 
$A_4^{(+8)}=(549,572)$, $A_5^{(+8)}=4$.
}
\label{table:yst_add}
\end{table*}

\section{Summary and discussion}
\label{sec:discuss}
In this short report, we investigated the influence of the degenerate self-loop 
 on attractor of the gene regulatory network model of the cell-cycle of  budding yeast, 


In the case of networks with degenerate self-loops removed from
 the original network $G^{(0)}$, 
only the point attractor appears because all of the self-loops are degenerate.
The attractor set of the network without the degenerate self-loops 
includes all attractors of the original network $G^{(0)}$.
 In addition, when self-regression loops and self-activation loops coexist, 
limit cycles with the period more than 2 appear other than point attractor, 
and many attractors of $G^{(0)}$ are not included in the attractor set, 
but the attractor with the largest basin size was relatively stable
against the deletions and additions of the self-loop.
Above result can apply to Boolean genetic network model of C.{\it elegans} early embryonic 
cell-cycle network as it is, because the self-loops of network are only 
self-inhibitation loops, and the attractors are 
only fixed points \cite{huang13,kinoshita18}. 

Note that necessary and sufficient condition 
that the network attractors does not become limit cycle but only 
 point attractors is not known yet \cite{tran13,goles13,richard15}.
However, 
we expect that the result in Subsec.\ref{subsec:removing} holds when 
at least  the attractors are only fixed points
in the random network with only degenerate self-loops.


There is a theorem in the graph theory \cite{goles13}:
{\it Consider a Boolean network 
such that each gene is governed with a threshold function. 
Then, if the associated incidence graph, without considering the diagonal elements,
is a directed acyclic graph (DAG) and the thresholds are non negative, $\theta_i \geq 0$,
then the attractors are only fixed points. }
The network of the budding yeast satisfies the following sufficient condition
for the fixed points.  
The result of  the Subsec.\ref{subsec:adding} 
seems to contradict above theorem at first glance.
However,  considering that the update rule  (\ref{eq:rule-2}) is different from 
one in Ref. \cite{goles13},
we can see that it is not necessarily contradictory to the theorem.






\appendix


\section{Case of  the random network models
}
\label{app:random}
\begin{figure}[htbp]
\begin{center}
\includegraphics[width=8.0cm]{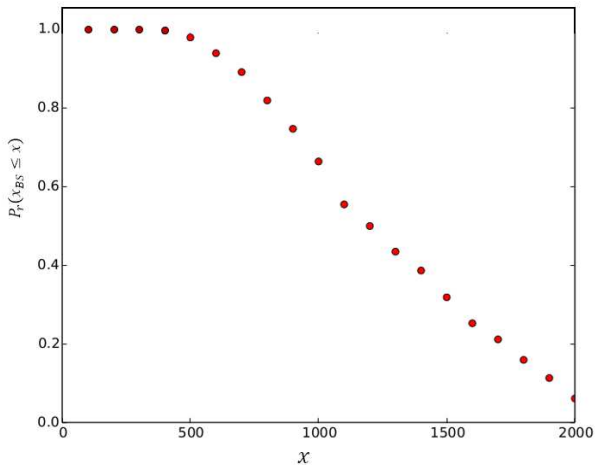}
  \caption{
The probability distribution  $Pr(x_{BS} \leq x)$  of 
the random network that the largest basin size $x_{BS}$ is 
smaller than $x$. 
We used 1000 network samples that 
the number of nodes is $11$, the number of active links is $14$, 
number of suppressing links is $15$, 
and (suppressing) self-loop number is $5$.
}
\label{fig:basin}
\end{center}
\end{figure}

We randomly construct the cell-cycle network with the same number of nodes and links
as the budding yeast, and examined the attractors and its basin size $x$
by re-linking in the network.
The number of nodes$=11$, the number of active links$=14$, 
number of suppressing links$=15$, and (suppressing) self loop number$=5$.
For each sample, attractors with the largest basin size were examined.
In the cases, all are point attractors because the networks satisfy 
the sufficient condition.
 Figure \ref{fig:basin} shows 
the probability distribution  $Pr(x_{BS} \leq x)$  of 
the random network that the largest basin size $x_{LBS}$ is 
smaller than $x$. 
It follows that  about 20 percent even on a random network maintaining 
the same structure as the budding yeast have attractor with 
the similar or the larger basin size  ($\geq 1700$) than the budding yeast.
The result is similar with those for  ES cell network of C.elegans in Ref.\cite{huang13}.



\end{document}